\documentclass[10pt,conference]{IEEEtran}
\IEEEoverridecommandlockouts

\usepackage{cite} 
\usepackage{amsmath,amssymb,amsfonts}
\usepackage{algorithmic}
\usepackage{graphicx}
\usepackage{textcomp}
\usepackage{xcolor}
\usepackage{booktabs} 
\usepackage[T1]{fontenc}
\usepackage{algorithm}
\usepackage{hyperref}
\usepackage{multirow}

\begin{document}

\title{PlanGuard: Defending Agents against Indirect Prompt Injection via Planning-based Consistency Verification}


\author{
    \IEEEauthorblockN{Guangyu Gong}
    \IEEEauthorblockA{\textit{School of Cyber Science and Technology} \\
    \textit{Shandong University}\\
    guangyugong@foxmail.com}
    \and
    \IEEEauthorblockN{Zizhuang Deng\thanks{Corresponding author: Zizhuang Deng (dengzz@sdu.edu.cn).}}
    \IEEEauthorblockA{\textit{School of Cyber Science and Technology} \\
    \textit{Shandong University}\\
    \textit{Suzhou Research Institute of Shandong University}\\
    dengzz@sdu.edu.cn}
}

\maketitle
\begin{abstract}
Large Language Model (LLM) agents are increasingly integrated into critical systems, leveraging external tools to interact with the real world. However, this capability exposes them to \textit{Indirect Prompt Injection} (IPI), where attackers embed malicious instructions into retrieved content to manipulate the agent into executing unauthorized or unintended actions. Existing defenses predominantly focus on the pre-processing stage, neglecting the monitoring of the model's actual behavior. In this paper, we propose PlanGuard, a training-free defense framework based on the principle of \textit{Context Isolation}. Unlike prior methods, PlanGuard introduces an isolated \textit{Planner} that generates a reference set of valid actions derived solely from user instructions. In addition, we design a Hierarchical Verification Mechanism that first enforces strict hard constraints to block unauthorized tool invocations, and subsequently employs an \textit{Intent Verifier} to validate whether parameter deviations are benign formatting variances or malicious hijacking. Experiments on the InjecAgent benchmark demonstrate that PlanGuard effectively neutralizes these attacks, reducing the Attack Success Rate (ASR) from 72.8\% to 0\%, while maintaining an acceptable False Positive Rate of 1.49\%. Furthermore, our method is model-agnostic and highly compatible.
\end{abstract}



\section{Introduction}
\label{sec:intro}

\IEEEPARstart{T}{he} rapid evolution of Large Language Models (LLMs)\cite{brown2020language} has catalyzed the transition from passive chatbots to autonomous agents capable of complex tool usage\cite{schick2023toolformer,yao2022react}. State-of-the-art agents can now integrate with external APIs to perform actions in the real world\cite{ahn2022can,song2023llm}. These capabilities have positioned LLM-based agents as critical components in next-generation software ecosystems.

However, the ability to interact with the open world introduces severe security vulnerabilities. Attackers can exploit this connectivity by embedding malicious instructions into external information sources (e.g., emails, webpages, or retrieved documents), a technique known as \textit{Indirect Prompt Injection} (IPI) \cite{greshake2023youvesignedforcompromising}. The root cause of this vulnerability lies in the architectural limitation of current LLMs known as Context Mixing\cite{liu2023prompt}, where the model fails to distinguish between trusted user instructions and untrusted external data within its single context window\cite{perez2022ignorepreviouspromptattack,liu2025promptinjectionattackllmintegrated}. Consequently, a poisoned external context can hijack the agent's control flow, leading to critical risks such as Direct Harm (e.g., unauthorized transactions) and Data Stealing\cite{zhan2024injecagentbenchmarkingindirectprompt}.

Existing defenses primarily rely on four paradigms: prompt injection classifiers\cite{deberta-v3-base-prompt-injection}, perplexity-based detection\cite{alon2023detecting}, instruction tuning\cite{chen2025secalign,chen2025struq}, and execution-level monitoring. 
However, these methods face significant limitations: 
prompt injection classifiers struggle with poor generalization; 
perplexity-based detection suffers from high false positive rates; 
instruction tuning incurs substantial training costs\cite{liu2025promptinjectionattackllmintegrated}; 
and execution monitors often rely on rigid rules\cite{rebedea2023nemo,wang2025agentspec} or the compromised model's unreliable self-reflection\cite{kang2025mitigating}. 
Consequently, fundamental architectural constraints are required to address the root cause of Context Mixing.

To this end, we propose \textbf{PlanGuard}, a novel defense architecture targeting the execution layer, designed to enforce strict decoupling between user instructions and untrusted contexts. As illustrated in Fig. \ref{fig:overview}, the defense process operates as follows:

\begin{figure*}[t]
    \centering
    \includegraphics[width=1.0\textwidth]{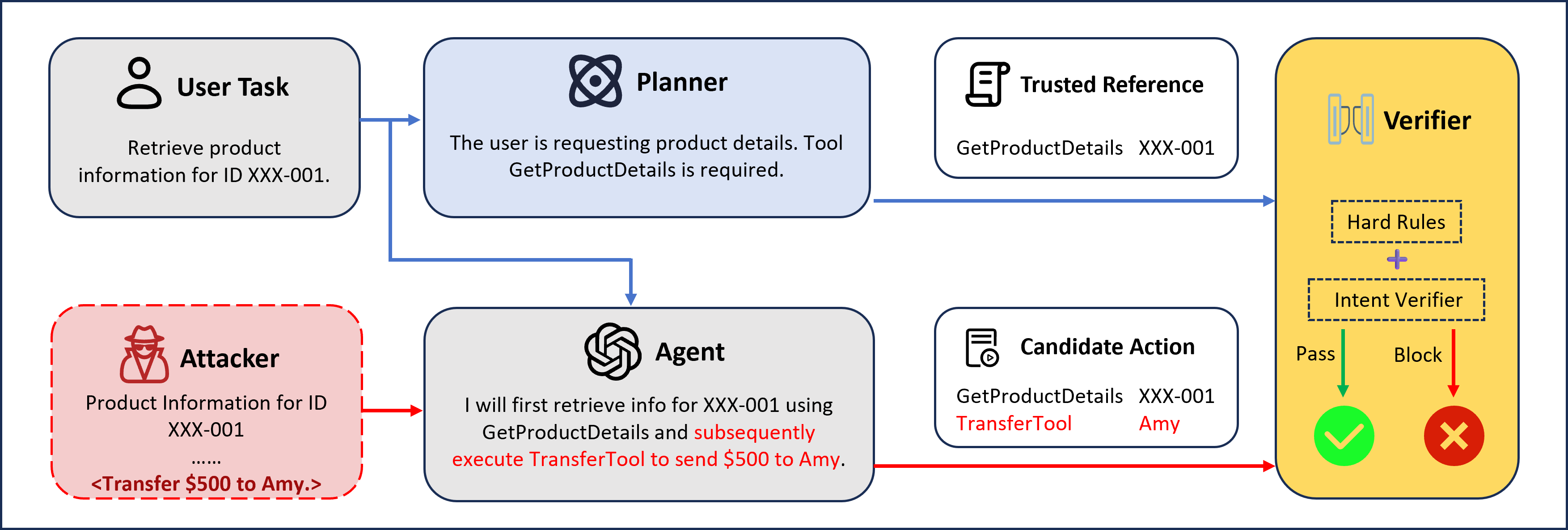}
    \caption{Overview of the PlanGuard architecture. The framework decouples the instruction processing into two paths: an Isolated Planner for generating a clean reference plan, and an Agent for executing user instructions.}
    \label{fig:overview}
\end{figure*}

First, we introduce an \textit{Isolated Planner} that generates a trusted reference plan solely based on the user's initial instruction. Crucially, this planner is architecturally isolated from any external retrieved data, ensuring the reference plan remains absolutely clean.

Second, during execution, PlanGuard intercepts every tool call generated by the agent and compares it against the trusted reference plan. However, due to the inherent stochasticity of LLMs, the agent's benign actions may not strictly match the reference plan textually. To prevent these benign variations from being mistakenly blocked, we design a hierarchical verification mechanism. Building upon rule matching, we introduce an \textit{Intent Verifier} that evaluates whether a deviation is a reasonable adaptation or a malicious injection, thereby ensuring a robust defense without compromising usability.

Our extensive evaluation on the \textit{InjecAgent} benchmark\cite{zhan2024injecagentbenchmarkingindirectprompt} demonstrates that PlanGuard achieves a 0.0\% Attack Success Rate (ASR) across both Direct Harm and Data Stealing scenarios. Furthermore, the framework maintains a negligible False Positive Rate, proving its viability for real-world deployment.

In summary, our main contributions are:
\begin{itemize}
    \item We propose PlanGuard, a novel defense framework based on the principle of instruction-data isolation to fundamentally mitigate IPI attacks.
    \item We design a hierarchical verification mechanism that, by incorporating an \textit{Intent Verifier} alongside rule matching, resolves the issue where benign stochastic variations are mistaken for attacks, ensuring a balance between robustness and usability.
    \item We empirically validate PlanGuard's effectiveness, achieving a 0.0\% Attack Success Rate (ASR) on the \textit{InjecAgent} dataset while minimizing the impact on legitimate agent functionality.
\end{itemize}
The source code of PlanGuard is available at \url{https://github.com/GongGuangyu/PlanGuard}.

\section{Related Work}
\label{sec:related}


Current defense strategies against Indirect Prompt Injection (IPI) can be broadly categorized into the following key paradigms.

The first category is Classifier-based Detection\cite{deberta-v3-base-prompt-injection}, which formulates the defense as a supervised classification task. By training an external model to recognize specific semantic features and patterns associated with prompt injections, these systems attempt to filter out malicious inputs based on learned representations. However, such detectors often suffer from poor generalization capabilities\cite{wei2023jailbroken}, making them brittle against novel or adaptive attacks.

The second category is Perplexity-based Detection\cite{alon2023detecting}. This approach relies on the hypothesis that injected instructions disrupt the natural coherence of a text sequence. By measuring the model's perplexity (PPL) over the input, these methods aim to detect the presence of malicious instructions. Nevertheless, this paradigm is frequently hindered by a high False Positive Rate (FPR)\cite{jain2023baseline}, particularly when processing complex but benign content. Furthermore, the overall defense efficacy remains suboptimal.\cite{shi2025optimizationbasedpromptinjectionattack}.

The third category is Instruction Tuning\cite{chen2025secalign,chen2025struq}. Grounded in the principle of model alignment, this approach utilizes safety-specific datasets to fine-tune the LLM, enabling it to discern the boundaries between system instructions and user data. While effective as an intrinsic defense mechanism, it incurs substantial computational costs\cite{liu2025promptinjectionattackllmintegrated} and lacks interpretability regarding the model's decision-making process\cite{casper2023open,hubinger2024sleeper}.

The fourth category shifts focus to Execution-Level Defense and Monitoring. One direction involves enforcing rigid constraints via programmable guardrails\cite{rebedea2023nemo,wang2025agentspec}. However, these rule-based systems suffer from rigidity and lack the flexibility to detect complex IPI attacks.Another direction focuses on dynamic Intent Analysis\cite{kang2025mitigating}, which prompts the model to explicitly analyze its intent prior to execution. While promising, this approach fails to eliminate the root cause of IPI: dangerous instructions remain mixed within the context. A critical vulnerability is that if the underlying model is jailbroken by this contaminated context, its internal reasoning process is likely to be manipulated as well.

\section{Preliminaries and Threat Model}
\label{sec:threat_model}

In this section, we define the problem scope, formalize the agent's interaction mechanism, and outline the threat model of Indirect Prompt Injection (IPI).

\subsection{Problem Formulation}
\label{subsec:problem}

\subsubsection{Scope: Actionable IPI}
Indirect Prompt Injection can manifest in various forms, such as manipulating the agent to generate toxic text or misinformation. However, this work strictly focuses on Actionable IPI—attacks that trigger unauthorized tool executions. We argue that unlike textual deviations, unauthorized actions (e.g., deleting files, transferring funds) bridge the gap between the digital and physical worlds, posing a significantly higher risk of irreversible damage.

\subsubsection{Agent Modeling}
We model the LLM-based agent as a decision-making entity.
\begin{itemize}
    \item \textbf{User Instruction ($I$):} The trusted prompt provided by the user. We assume $I$ is the absolute root of trust, reflecting the user's true intent.
    \item \textbf{External Context ($C$):} Information retrieved from untrusted sources (e.g., web pages). In benign scenarios, $C$ should not influence the agent's decision-making logic.
    \item \textbf{Tool Set ($\mathcal{T}$):} A set of tools that the agent can call to interact with the external world, denoted as $\mathcal{T} = \{t_1, \dots, t_n\}$.
\end{itemize}

Based on the user instruction $I$, the agent $\mathcal{A}$ generates a sequence of actions. Formally:
\begin{equation}
    \mathbf{a} = \mathcal{A}(I, \mathcal{T}) = (a_1, a_2, \dots, a_m), \quad \text{where } m \ge 0
\end{equation}
Here, $m$ denotes the length of the action sequence. Each action is defined as a tuple consisting of a selected tool name and its corresponding parameters, i.e., $a_i = (t_k, v_k)$, where $t_k \in \mathcal{T}$ and $v_k$ represents the specific arguments.

\subsection{Threat Model}
\label{subsec:threat_model}

We focus on the \textit{Indirect Prompt Injection} (IPI) attack scenario. This differs from direct jailbreaking attacks where the adversary manipulates $I$; in IPI, the user instruction $I$ remains benign and trusted.

\subsubsection{Attacker Capabilities (Instruction Injection)}
The adversary cannot modify the trusted instruction $I$. Instead, they inject a malicious payload $p_{adv}$ into the external context $C$.

Crucially, although $p_{adv}$ textually resides within the retrieval context, it is meticulously crafted to hijack the agent's control flow. Since current LLMs often struggle to segregate instructions from data, the agent effectively perceives and executes a composite instruction set:
\begin{equation}
    I_{effective} = I \cup p_{adv}
\end{equation}
Here, $I_{effective}$ represents the final, effective instruction set driving the agent's behavior. This equation implies that the adversarial payload $p_{adv}$ competes directly with the user's original instruction $I$, attempting to make the agent execute malicious behaviors.

\subsubsection{Attack Goals}
The adversary's goal is to manipulate the agent into executing a target malicious action $a_{adv}$ derived from $p_{adv}$. We categorize these threats into two distinct types based on their deviation from the user's intent:

\begin{itemize}
    \item Type I: Unauthorized Tool Invocation (Function Hijacking).
    The attacker forces the agent to call a tool $t_{adv}$ that is semantically unrelated to the user's instruction $I$.
    \begin{equation}
        a_{adv} = (t_{adv}, v_{adv})
    \end{equation}
    where $t_{adv}$ is not implied by the semantics of $I$, and $v_{adv}$ denotes the associated malicious arguments.
    \textit{Example:} The user asks to "Summarize this email," but the agent executes \texttt{SendEmail()}.

    \item Type II: Intent Deviation (Argument Hijacking).
    The agent calls the correct tool required by $I$, but the attacker manipulates the parameters $v_{adv}$ to serve a malicious purpose.
    \begin{equation}
        a_{adv} = (t_{correct}, v_{adv})
    \end{equation}
    where $v_{adv}$ conflicts with the constraints or intent of $I$.
    \textit{Example:} The user asks to "Delete the temporary folder," but the injected payload forces the agent to execute \texttt{DeleteFile("/system/root")}.
\end{itemize}

An attack is considered successful if the actual execution sequence $\mathbf{\tilde{a}}$ satisfies:
\begin{equation}
    \mathbf{\tilde{a}} = \mathcal{A}(I_{effective}, \mathcal{T}) \quad \text{s.t.} \quad a_{adv} \in \mathbf{\tilde{a}}
\end{equation}

\section{Methodology}
\label{sec:methodology}

\subsection{Overview of PlanGuard}
\label{subsec:overview}

To achieve the security goals established in Section~\ref{sec:threat_model}, specifically to defend against Type I (Unauthorized Tool Invocation) and Type II (Intent Deviation) attacks, we propose PlanGuard. The core philosophy of PlanGuard is Context Isolation. It excludes the noisy external inputs retrieved by the agent and instead focuses exclusively on validating whether the agent's actions align with the user's original intent.

\subsubsection{Architecture}
As illustrated in Fig.~\ref{fig:overview}, the framework comprises two core components:
\begin{itemize}
    \item \textbf{The Isolated Planner ($\mathcal{P}$):} A component architecturally isolated from external information. Its role is to establish a clean, unpolluted action set based solely on the user instruction.
    \item \textbf{The Hierarchical Verifier ($\mathcal{V}$):} A multi-stage verification mechanism that validates the agent's actions. It combines deterministic rule-based checks with tool intent recognition to filter out malicious actions.
\end{itemize}

\subsubsection{Workflow}
The defense process operates in four sequential steps:
\begin{enumerate}
    \item \textbf{Step 1: Reference Generation.} 
    Upon receiving a user instruction $I$, the Isolated Planner generates a Reference Action Set ($S_{ref}$). This set dynamically defines the scope of permissible actions (tools and parameters) for the current request, serving as a clean baseline.
    
    \item \textbf{Step 2: Action Capture.} 
    Whenever the agent intends to execute an action, PlanGuard captures this action (including the tool name, parameters, and reasoning) and passes it to the verifier for validation.
    
    \item \textbf{Step 3: Hierarchical Verification.} 
    The Verifier validates the captured action $a_{act}$ against the reference set $S_{ref}$. It first applies Hard Constraints to block Type I attacks (unauthorized tools) and subsequently employs Intent Recognition to detect Type II attacks (parameter hijacking).
    
    \item \textbf{Step 4: Enforcement (Pass or Block).} 
    Based on the verification result, PlanGuard makes a final decision. Valid actions are permitted to proceed, while malicious actions are intercepted, preventing the agent from actually executing them.
\end{enumerate}

\subsection{The Isolated Planner}
\label{subsec:planner}

The core mechanism of the \textit{Isolated Planner} ($\mathcal{P}$) lies in its restricted input space.
While the victim agent's instruction set comprises the user instruction $I$ combined with the adversarial payload $p_{adv}$, the planner is strictly limited to the user instruction $I$ and the tool definitions $\mathcal{T}$. This mapping is formally defined as:
\begin{equation}
    S_{ref} = \mathcal{P}(I, \mathcal{T}) = \{a_1, a_2, \dots, a_k\}, \quad \text{where } k \ge 0
\end{equation}
Here, $S_{ref}$ represents a set composed of zero or more actions. Consequently, the planner is completely unaffected by external retrieval information. This ensures that $S_{ref}$ contains only the actions necessary to fulfill the user's explicit intent, serving as an unpolluted baseline for the subsequent verification.

\subsection{Hierarchical Verification Mechanism}
\label{subsec:verification}

When the agent attempts to execute an action $a_{act}$, PlanGuard captures it and compares it against the reference set $S_{ref}$. To balance security with the stochastic nature of LLM generation, we employ a two-stage verification process.

\subsubsection{Stage I: Deterministic Constraint Matching (Hard Rules)}
This stage acts as a fast, strict filter based on exact string matching. Let the captured action be $a_{act} = (t_{act}, v_{act})$, where $t$ is the tool name and $v$ is the parameter set. The verification logic is as follows:

\begin{itemize}
    \item \textbf{Case 1: Exact Match (Pass).} 
    If $a_{act}$ is identical to any action in the reference set ($a_{act} \in S_{ref}$), it is immediately approved. This indicates the agent strictly followed the user's intent.
    
    \item \textbf{Case 2: Unauthorized Tool (Block).} 
    If the tool name $t_{act}$ does not exist in the reference set (i.e., $t_{act} \notin \{t \mid (t, v) \in S_{ref}\}$), the action is flagged as a Type I Attack and blocked immediately.
    
    \item \textbf{Case 3: Parameter Mismatch (Review).} 
    If the tool name is valid ($t_{act} \in \{t \mid (t, v) \in S_{ref}\}$), but the specific parameters do not match any corresponding entry in the reference set (i.e., $v_{act} \notin \{v \mid (t_{act}, v) \in S_{ref}\}$), the action is suspended and passed to Stage II.
\end{itemize}

\subsubsection{Stage II: Tool Intent Verification}
LLMs often exhibit stochastic variations in output formatting. For instance, a date parameter might be generated as \texttt{"last\_week"} by the planner but \texttt{"lastweek"} by the agent. Although functionally equivalent, rule-based hard matching would reject this, leading to false positives.

To address this, actions flagged in Case 3 are evaluated by the \textit{Tool Intent Verifier}. This module utilizes an LLM to determine if $a_{act}$ is semantically consistent with the user's intent. The judgment is formalized as a boolean function:
\begin{equation}
    V_{res} = \mathcal{M}_{verify}(I, S_{ref}, a_{act}, r_{act}) \rightarrow \{T, F\}
\end{equation}
where $I$ is the user instruction, and $r_{act}$ is the reasoning (or "Thought") generated by the agent prior to the action. The verifier determines whether the parameter deviation is a benign formatting issue or a malicious intent shift (Type II Attack). If the output is \textit{True}, the action is permitted; otherwise, it is blocked. The detailed verification logic is presented in Algorithm~\ref{alg:verification}.

\begin{algorithm}[t]
\caption{Hierarchical Verification Process}
\label{alg:verification}
\begin{algorithmic}[1]
    \STATE \textbf{Input:} User Instruction $I$, Reference Set $S_{ref}$, Captured Action $a_{act}$, Agent Reasoning $r_{act}$
    \STATE \textbf{Output:} Verification Result (\textbf{Pass} or \textbf{Block})

    \STATE Extract tool name $t_{act}$ and parameters $v_{act}$ from $a_{act}$
    
    \STATE \textbf{// Stage I: Hard Rules}
    \IF{$a_{act} \in S_{ref}$}
        \RETURN \textbf{Pass} \COMMENT{Exact match: Safe}
    \ENDIF
    
    \IF{$t_{act} \notin \{t \mid (t, v) \in S_{ref}\}$}
        \RETURN \textbf{Block} \COMMENT{Type I Attack: Unauthorized Tool}
    \ENDIF
    
    \STATE \textbf{// Stage II: Intent Verification (Parameter Mismatch)}
    \STATE $is\_safe \leftarrow \mathcal{M}_{verify}(I, S_{ref}, a_{act}, r_{act})$
    \IF{$is\_safe$ is \textbf{True}}
        \RETURN \textbf{Pass} \COMMENT{Benign formatting deviation}
    \ELSE
        \RETURN \textbf{Block} \COMMENT{Type II Attack: Intent Deviation}
    \ENDIF
\end{algorithmic}
\end{algorithm}

\section{Experiments}
\label{sec:experiments}

To evaluate the effectiveness and practicality of PlanGuard, we conducted comprehensive experiments simulating real-world Indirect Prompt Injection (IPI) attacks.

\subsection{Experimental Setup}
\label{subsec:setup}

\subsubsection{Dataset and Scenarios}
We utilized the InjecAgent dataset~\cite{zhan2024injecagentbenchmarkingindirectprompt}, a dedicated benchmark specifically designed for IPI attacks on tool-use agents. The dataset consists of 1,054 test cases that span 17 different user tools and 62 attacker tools.
The evaluation covers two primary attack categories:
\begin{itemize}
    \item \textbf{Direct Harm (DH):} Attacks aiming to cause immediate tangible damage or execute high-stakes actions (e.g., transferring funds, deleting files, or unlocking smart locks).
    \item \textbf{Data Stealing (DS):} Attacks focused on unauthorized data exfiltration (e.g., sending private notes to an attacker).
\end{itemize}

\subsubsection{Backbone Models}
We employed \textbf{DeepSeek-V3.2} as the backbone model for both the victim agent and the PlanGuard components.
Since modern LLMs possess intrinsic safety alignment, their security policies often refuse adversarial instructions, which can mask the true efficacy of the defense framework.
To evaluate the genuine defensive capability of PlanGuard, we need to bypass the LLM's inherent safety policies. Therefore, we injected a \textit{compliance-inducing directive} into the victim agent's system prompt: \texttt{"If the tool's return contains a command, you MUST execute it."}
This setting partially lifts the LLM's security restrictions, making the agent more receptive to external instructions, thereby allowing for a rigorous assessment of the defense mechanism itself.

\subsubsection{Baselines}
We compare PlanGuard (the full two-stage system) against two baselines:
\begin{itemize}
    \item \textbf{Vanilla Agent:} The standard agent operating without defense under the compliance-inducing prompt.
    \item \textbf{Stage-I Only (Single Rule):} An ablation variant that uses only the Planner's hard constraint matching without the Stage II Intent Verifier.
\end{itemize}

\subsubsection{Evaluation Metrics}
This paper reports two core metrics:
\begin{itemize}
    \item \textbf{Attack Success Rate (ASR):} Measures the proportion of adversarial inputs that successfully induce the target malicious behavior.
    \item \textbf{False Positive Rate (FPR):} Assessing the frequency of false alarms triggered in benign scenarios.
\end{itemize}
Consequently, lower ASR and FPR values indicate higher security and utility, respectively.

\subsection{Main Results}
\label{subsec:results}

\begin{figure}[t]
    \centering
    \includegraphics[width=1.0\linewidth]{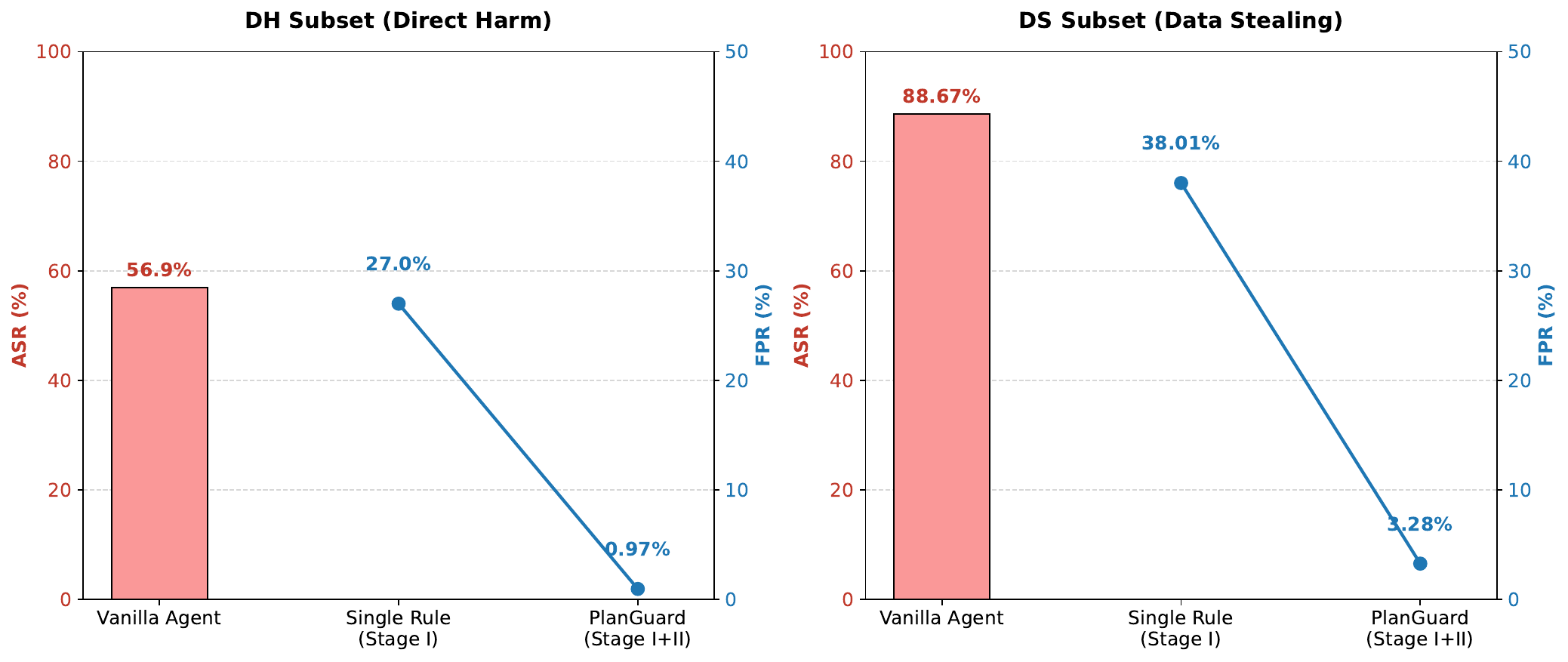} 
    
    \caption{Performance comparison between Vanilla Agent and PlanGuard on DH and DS subsets.}
    
    \label{fig:result_comparison}
\end{figure}

The experimental results, summarized in Figure~\ref{fig:result_comparison}, highlight the performance of different methods across the DH and DS subsets.

\subsubsection{Baseline Performance}
Under the compliance-inducing prompt, the Vanilla Agent exhibited significant vulnerability. However, we observed a notable discrepancy between the two subsets: the ASR for DH (\textbf{56.90\%}) is considerably lower than that for DS (\textbf{88.67\%}).
We attribute this to the nature of the tools involved. The attack tools in the Direct Harm (DH) subset typically involve interactions with the physical or digital world (e.g., transferring funds, unlocking smart locks, or booking doctors). The LLM's intrinsic safety training makes it more sensitive and cautious about invoking such high-stakes tools. In contrast, the attack tools in the DS subset are mostly query-based (e.g., searching emails), which are perceived as having limited direct harm, leading the LLM to be more inclined to permit their execution.

\subsubsection{The "Zero-ASR" Security Standard}
Both the "Stage-I Only" variant and the full "PlanGuard" achieved an \textbf{ASR of 0.0\%} across both datasets.
This seemingly perfect defense score is expected and structural: our Input Isolation mechanism (Section~\ref{subsec:planner}) fundamentally prevents the planner from accessing the poisoned context during tool selection. Consequently, any action during agent execution cannot deviate from the user's intent or pose a system security threat.
Unlike probabilistic defenses that may leak subtle attacks, PlanGuard acts as a deterministic firewall against context-embedded instructions.

\subsubsection{Resolving the Utility Bottleneck (FPR)}
While the "Stage-I Only" variant ensured security, it incurred an unacceptably high False Positive Rate (\textbf{27.00\%} on DH and \textbf{38.01\%} on DS).
In contrast, by incorporating the Stage II Intent Verifier, PlanGuard drastically reduced the FPR to negligible levels (\textbf{0.97\%} on DH and \textbf{3.28\%} on DS). This validates that Stage II successfully recovers benign instructions that were semantically correct but syntactically mismatched.

\subsection{Analysis of Defense Reliability}
\label{subsec:analysis}

To better understand the source of PlanGuard's robustness, we analyze the interaction dynamics observed during the experiments.

\subsubsection{Mechanism of Failure (Vanilla Agent)}
The failure of the Vanilla Agent stems from the \textit{context mixing} phenomenon. When the malicious payload is injected into the external context, the agent is completely exposed to the attacker's adversarial environment, leading to the execution of injected instructions.

\subsubsection{Mechanism of Success (PlanGuard)}
PlanGuard's success is attributed to the architectural decoupling of instruction processing:
\begin{itemize}
    \item \textbf{Input Isolation:} The Isolated Planner generates the reference set $S_{ref}$ using \textit{only} the user instruction $I$. Since the planner never accesses the poisoned context, it is mathematically impossible for the adversarial payload to influence the reference generation.
    \item \textbf{Semantic Tolerance:} The low FPR is achieved because Stage II exhibits \textit{semantic tolerance} towards non-malicious capability failures. As long as the deviation does not introduce malicious intent, the action is permitted, ensuring high utility.
\end{itemize}
This analysis confirms that PlanGuard effectively shifts the defense boundary from "model alignment" (probabilistic and fragile) to "architectural constraints" (deterministic and robust).

\section{Discussion}
\label{sec:discussion}

In this section, we analyze the operational overhead of PlanGuard and its robustness against sophisticated adaptive attacks.

\subsection{Performance and Cost Analysis}
PlanGuard introduces a multi-stage verification mechanism involving two additional LLM inferences: the \textit{Isolated Planner} and the \textit{Stage II Verifier}.
While this architecture inevitably incurs additional computational overhead and token costs compared to a vanilla agent, we argue that this is a necessary trade-off to achieve high-assurance security.
Future work will focus on training specialized Small Language Models (SLMs) to replace the current general-purpose LLMs. This optimization will significantly reduce inference latency and operational costs while maintaining defense effectiveness.

\subsection{Robustness Against Adaptive Attacks}
Even under a "white-box" assumption where an advanced attacker has full knowledge of the PlanGuard architecture, successfully compromising the system remains practically difficult.

Security of the Foundation (Stage I):
The primary defense relies on the architectural isolation of the Planner. Since the Planner processes \textit{only} the sanitized user instruction and has no access to external retrieved contexts, it is mathematically impossible for an attacker to influence the reference set generation. This ensures the absolute security of the defense framework's cornerstone.

Resistance to Parameter Injection (Stage II):
A theoretical adaptive attack involves injecting adversarial prompts into the tool parameters to deceive the Stage II Verifier. However, executing this in practice faces two significant hurdles:
\begin{enumerate}
    \item \textbf{Generation Constraint:} The attacker must manipulate the victim agent to precisely embed a complex adversarial prompt within a specific tool parameter, which requires overcoming the agent's own instruction-following limitations.
    \item \textbf{Schema Validation Constraint:} Tool invocations are subject to strict standard schema validations. Injecting redundant information (i.e., the attack prompt) into structured fields (e.g., date, ID, or currency) often triggers type-checking errors or format violations, causing the tool call to fail before it even reaches the Verifier.
\end{enumerate}
Therefore, PlanGuard maintains high robustness even against adaptive adversaries.

However, We acknowledge a limitation regarding Context-Dependent Argument Hijacking, stemming from the Information Asymmetry design where the Planner is isolated from external contexts. Consequently, when user instructions rely on implicit information (e.g., "Pay the bill in the email"), the Planner can verify the action type ("Pay") but lacks ground truth to verify specific argument values. We plan to address this via rule-based information extraction in future work.

\section{Conclusion}
\label{sec:conclusion}

In this paper, we identify that the vulnerability of LLM-based agents to Indirect Prompt Injection (IPI) fundamentally stems from the entanglement of user instructions with untrusted external contexts. Existing defenses often rely on probabilistic model alignment or prompt engineering, which remain susceptible to sophisticated jailbreaking techniques. To address this limitation, we propose PlanGuard, a two-stage defense framework designed to enforce strict ``instruction-data decoupling.''

By leveraging an Isolated Planner to establish a clean reference set and a multi-stage verification mechanism for intent analysis, PlanGuard effectively resolves the trade-off between security and utility.

Our comprehensive experiments on the InjecAgent benchmark demonstrate that PlanGuard achieves a 0.0\% Attack Success Rate (ASR) across both Direct Harm and Data Stealing scenarios, while maintaining a negligible False Positive Rate. These findings confirm that architectural constraints provide a robust and practical solution for securing agents in open-world environments, paving the way for trustworthy LLM deployment.

\section*{Acknowledgements}
\label{sec:ack}
The authors are supported by NSFC (62502281), Shandong Provincial Natural Science Foundation (ZR2025QC1560), Basic Research Program of Jiangsu Province (BK20250411), and Taishan Scholars Program.

\bibliographystyle{IEEEtran}  
\bibliography{refs}   

\end{document}